\documentclass[a4paper,11pt]{article}
\usepackage[utf8]{inputenc}
\usepackage[utf8]{inputenc}
\usepackage[english]{babel}
\usepackage[T1]{fontenc}
\usepackage{amsmath}
\usepackage{graphicx}
\usepackage{subcaption}
\usepackage{authblk}

\usepackage[round]{natbib} 
\bibpunct{(}{)}{;}{a}{}{,}

\title{An analysis of the vectorial capacity using moment--generating functions}
\author{Daniel A.M. Villela\\
{\tt dvillela@fiocruz.br}\\
Programa de Computação Científica (PROCC)\\
Fundação Oswaldo Cruz (FIOCRUZ/Brasil)}
\date{}

\begin{document}

\maketitle

{\bf Abstract}

This paper describes a technique for analyzing the stochastic structure of the vectorial capacity using moment--generating functions. In such formulation, for an infectious disease transmitted by a vector, we obtain the generating function for the distribution of the number of infectious contacts (e.g., infectious bites from mosquitoes) between vectors and humans after a contact to a single infected individual.
This approach permits us to generally derive the moments of the distribution and, under some conditions, derive the distribution function of the vectorial capacity. 
A stochastic modeling framework is helpful for analyzing the dynamics of disease spreading, such as performing sensitivity analysis.

\section*{Introduction}

The field of mathematical modeling of infectious diseases had seminal works by 
\cite{ross1910prevention} and \cite{macdonald1957epidemiology} that considered the dynamics of malaria spreading.
The works by \cite{garrett1964human,garrett1964prognosis,garrett1969malaria}
introduced the concept of Garret--Jones' vectorial capacity, which is the number of future inoculations that arise from an infectious case. Such modeling framework has been used not only for malaria, but also for other vector--transmitted diseases.  This is the case of dengue, a disease already well-known for causing epidemics in various parts of the world and also emerging infections, such as chikungunya. 

The modeling framework has been described in the literature, for instance by \cite{anderson1991infectious}.
More recent works also revisited the issue in a tutorial--oriented style or a historical perspective.
\cite{smith2004statics} revisit the formulae involved in the statics and the dynamics of malaria infections.
\cite{smith2012ross} provide the history behind the contributions of Ross and MacDonald. \cite{massad2012vectorial} provide a treatment that explains the dimensionless property of the number.

\cite{bailey1982biomathematics,bailey1975mathematical} analyzed epidemics using stochastic models, including using moment--generating functions, which provided a seminal body of work in its own right, but did not model the vectorial capacity as presented here.  Also, for a general epidemic process, for instance, \cite{ball1995strong} presents approximations results, using moment--generating functions.  

Nevertheless,
the treatment presented here using moment--generating functions to describe the vectorial capacity has not appeared elsewhere, to the best of our knowledge, and can be quite important to advance the study of the vectorial capacity.


\section*{Model}

In order to describe the vectorial capacity we need to describe the biting process and the survival of vectors. Vectors are species that can transmit pathogens causing diseases.  In the case of malaria or dengue, vectors are typically mosquitoes.  For this reason, we call the process between the vector and humans as a biting process, but it could be applied to species, other than mosquitoes.

The total number of vectors is a random variable $M$ and the number of humans is assumed fixed at $h$ individuals. 

\subsection*{ Moment generating functions}

For any generic random variable $X$ the moment--generating function (MGF) is defined by $X^{*}(\theta)= E[e^{\theta X}]$.\footnote{The Laplace transform is the case when there is a variable replacement from $\theta$ to $-s$, i.e. $E[e^{-s X}]$.}
Therefore,
\begin{equation}
X^{*}(\theta) = \int_0^\infty e^{\theta x} P(X=x) dx.
\end{equation}


For discrete distributions, which is the case of the variable that describes the vectorial capacity is convenient to work with the z--transform:
\begin{eqnarray*}
X^{*}(z) & = & \sum_{k=0}^\infty z^k P(X=k), z= e^\theta,  \mbox{if X is discrete}.
\end{eqnarray*}


\subsection*{The biting process}

The individual number of bites is described by a random variable $B$ that has distribution $B(x)$ and MGF $B^{*}(z)$.
When counting the multiple individuals, with the assumption that the count numbers are i.i.d, we have a total number $B_T = \sum_{i=0}^M B(z)$.
Hence, the MGF for $B_T$ is $B_T^{*}(z) = M(B(z))$.

\subsection*{The survival process}

Let $S$ be a random variable that describes the survival time of a female mosquito after biting an infected person and $S_v$ the survival time after the incubation period. The distribution of $S$ and $S_v$ are given by functions $S(\tau)$
and $S_v(\tau)$, respectively.  The moment generating functions follow from the definition $S(\theta)=E[e^{\theta S}]$ and $S_v(\theta)=E[e^{\theta S_v}]$.

\subsection*{Biting as a Poisson process}

We now introduce an initial assumption that intervals between contacts for any vector are independent identically distributed random variables.  A second assumption is that the distribution of the intervals between contacts is exponential with mean $\lambda$.

Let $B$ be a random variable that describes the number of bites by a mosquito in a given period of time.  Since the intervals between contacts are exponentially distributed, then $B$ has a Poisson distribution.
Therefore, the number $B$ of bites (individual) over a period $t$ has a density distribution given by
\[ P(B = k) =  e^{-\lambda t} \frac{(\lambda t)^k}{k !} \]
and its MGF given by
\begin{equation}
\label{eq:momentB}
B^{*} (z) = e^{-\lambda t (1-z)}.
\end{equation}

Taking into account all mosquitoes, then the MGF of the total number of bites is the product of moment generating function of $B$ expressed in Eq. \ref{eq:momentB}:
\begin{equation}
B_T^{*} (z) = M(B^{*}(z))
\end{equation}
where $M(z)$ is the moment generating function of number of mosquitoes.  For instance, if the total number of mosquitoes is constant at $M=m$:
\begin{equation}
\label{eq:BT}
B_T^{*} (z) = e^{-m \lambda t (1-z)}.
\end{equation}

Now we turn to the number $B_L$ of bites of a mosquito during its virus--carrying state. 
The number $B_L$ of bites in infective state is also described by a Poisson distribution with parameter $\lambda$.
%
If we condition the number of bites to the time $S$, we have:
\[ \{B_L |  S=t \} = B \sim Poisson(\lambda t).\] 

Following Eq. \ref{eq:momentB}, the moment--generating function is then given by:
\begin{equation}
B_L^{*}(z | S=t ) = e^{-\lambda t (1-z)}
\end{equation}

We find $B_L^{*}$ by integrating $B_L^{*}(z | S=t)$ over the spectrum of the probability density function of $S$:
\begin{equation}
\label{eq:BL}
B_L^{*}(z) = \int_0^{\infty} e^{-\lambda t (1-z)} P(S=t) dt = S^{*}(\lambda(z-1))
\end{equation}

\subsection*{Vectorial capacity process}

We start with a single infected human.
Let $N_S$ be a random variable that describes the number of mosquitoes that bite this infected individual, get blood with virus and also survive past the incubation period.  Then the total number $B_{SI}$ of bites given by this group is the sum of bites per individual, i.e., $B_{SI} = \sum_{j=1}^{N_s} B_{L,j}$.



Therefore, assuming all $B_{L,j}$ are i.i.d. (independent, identically distributed), and using Eq. \ref{eq:BL}, the moment--generating function is
\begin{equation}
B_{SI}^{*}(z) = N_s(S^{*}(\lambda(z-1)))
\end{equation}

Now we have to find the moment generating function of $N_S$.
Let the incubation time of the pathogen be a random variable $V$.

There are a number of bites $B_I$ that effectively infect mosquitoes.  Some of these will not survive past the incubation time $V$.  Therefore, there is a probability that $S \ge V$.  From all bites that infect mosquitoes, the number $N_S$ is given then by a binomial distribution:
%
\[ N_S | V, B_I, S \sim Binom(B_I, P(S \ge V)) \]
where the probability $p_V$ that a vector survives past the incubation period is given by 
\[ p_V=P(S \ge V) = \int_0^{\infty} P(S \ge t)P(V=t) dt.\]

The moment--generating function for $N_S$ conditioned on $B_I$, given its binomial distribution, is 
\begin{equation}
N_S^{*} (z | B_I=k) = (p_V z + 1 - p_V)^{k}
\end{equation}

The MGF of the number of mosquitoes that reach the state of effective vectors (past the incubation period), conditioned on the total number of virus--carrying ones, is given by
\begin{equation}
\label{eq:BSI}
B_{SI}^{*}(z | B_I=k) = N_s(S^{*}(\lambda(z-1))) = (p_V (S^{*}(\lambda(z-1))) + 1 - p_V)^{k}.
\end{equation}

Now the number $B_I$ of bites that infect mosquitoes has a binomial distribution, and its parameter is the probability $q$ to bite the single human infected.  If we assume that the probability to bite the single human infected is the same (uniformly distributed) across mosquitoes, then $q = h_i/h$, as function of the number $h_i$ of infected humans.  If $h_i=1$, we have simply $q=1/h$.
\[ B_I \sim Binom (B_T, q). \]

Therefore, the moment--generating function of $B_I$ can be expressed from the moment--generating function of $B_T$ derived in Eq. \ref{eq:BT}:
\begin{equation}
B_I^{*} (z) = B_T^{*}(q z + 1- q) = e^{-m \lambda (1 - (q z + 1-q) )}
\end{equation}

Finally, using this MGF of $B_I^{*}(z)$ and the result in Eq. \ref{eq:BSI}:
\begin{eqnarray}
\label{eq:general}
B_{SI}^{*}(z ) & = & B_I^{*} (p_V (S^{*}(\lambda(z-1))) + 1 - p_V) \nonumber \\ 
& = & B_T^{*}(q (p_V(S^{*}(\lambda(z-1))) + 1 - p_V) + 1-q) \nonumber \\
B_{SI}^{*}(z) & = & M^{*}(B^{*}(q (p_V(S^{*}(\lambda(z-1))) + 1 - p_V) + 1-q)) 
\end{eqnarray}

Eq. \ref{eq:general} is a more general result to derive the genrating function.
In particular, under the assumption of a Poisson distribution for the number of bites and a fixed number $m$ of mosquitoes:
\begin{eqnarray}
\label{eq:moment}
 B_{SI}^{*}(z ) & = & e^{ -m \lambda (1 - (q (p_V (S^{*}(\lambda(z-1))) + 1 - p_V) + 1- q))  }
\end{eqnarray}

\subsection*{Lifetime as an exponentially--distributed random variable}

We now assume that lifetime (or at least adult lifetime) is exponentially distributed.  Then, upon this assumption, $S$ also has an exponential distribution.
The probability to survive a given period of time, say a day, is $p= e^{-g}$ and for $t$ days we have $p^t = e^{-g t}$

Therefore, we can express $P(S>V)$ in terms of the moment--generating function of $V$:
\begin{equation}
p_V = P(S \ge V) = \int_0^{\infty} e^{- g t } P(V=t) dt = V^{*}(\theta)|_{\theta=-g}
\end{equation}

\subsection*{Moments of the vectorial process}

If we consider the latent period to be deterministic, given by $v$ days,
then $V^{*} (z) = z^v$ (discrete case) or $V^{*}(\theta) = e^{\theta v}$ (continuous case).

Also, the lifetime distribution with an exponential distribution has a probability density function $P(S=t) = g e^{-gt}$ and its MGF
$S^{*}(\theta) = -g/(\theta-g)$.

Then we apply $S^{*}(\theta) = -g/(\theta-g)$ and $p_V=e^{-g v}$ into Eq. \ref{eq:moment}:
\begin{eqnarray}
\label{eq:momentspec}
B_{SI}^{*} (z) 
& = & e^{ -m \lambda \frac{e^{-gv}}{h} \left(1 - \frac{-g}{\lambda(z-1)-g}\right) }
\end{eqnarray}

If we take the first derivative of $B_{SI}^{*}(z)$, for $z=1$, we find the mean $E[B_{SI}]$:
\begin{equation}
\frac{d B_{SI}^{*}(z)}{dz}|_{z=1} = E[B_{SI}] = \frac{m \lambda^2 e^{-g v}}{h g}
\end{equation}
which is the well-known formula for vectorial capacity.

If we take the second--order derivative, for $z=1$, we find $E[B_{SI}^2] - E[B_{SI}]$.
\begin{eqnarray}
 \frac{d^2 B^{*}_{SI} (z)}{dz^2} & = & \frac{d B_{SI}^{*}}{dz} (z) \frac{m \lambda^2 e^{-g v} g}{h (\lambda(z-1)-g)^2}  + B_{SI}^{*} (z) \frac{m \lambda^2 e^{-g v} g}{h}\frac{2\lambda}{(\lambda(z-1)-g)^3} \nonumber \\
 \frac{d^2 B^{*}_{SI}}{dz^2} |_{z=1} & = & (E[B_{SI}])^2 + \frac{2m \lambda^3 e^{-g v}}{h g^2} = (E[B_{SI}])^2 + E[B_{SI}] \frac{2\lambda}{g}. \nonumber
\end{eqnarray}

Thus, the variance is found:
$Var(B_{SI}) = E[B_{SI}](1 + \frac{2 \lambda}{g}) = \frac{m \lambda^2 e^{-g v}}{h g} (1 + \frac{2 \lambda}{g})$.

\subsection*{Inverse transform to obtain the distribution}

Using the power series $e^x = \sum_{k=0}^{\infty} x^k/k!$,
we rewrite Eq. \ref{eq:momentspec}: 
\begin{equation*}
B_{SI}^{*} (z) = e^{-m \lambda \frac{e^{-gv}}{h}} \sum_{k=0}^{\infty} [f(z)]^k/k!
\end{equation*}
where $f(z) = m \lambda \frac{e^{-g v}}{h} \left(\frac{g}{g+ \lambda(1-z)}\right)$.
Rewriting previous equation after some algebraic modifications,
\begin{equation}
B_{SI}^{*} (z) = e^{-m \lambda \frac{e^{-g v}}{h}} \sum_{k=0}^{\infty} \frac{1}{k!} \left(\frac{m \lambda e^{-gv}g}{h(g+\lambda)}\right)^k \frac{1}{(1 -\frac{\lambda}{g+\lambda} z)^k}
\end{equation}

Taking the inverse z-transform we find the probability density function for $B_{SI}$:
\begin{equation}
P(B_{SI}=n) = e^{-m \lambda \frac{e^{-g v}}{h}} \left( u_n + \left(\frac{\lambda}{g+\lambda}\right)^n \sum_{k=1}^{\infty} {\binom{n+k-1}{k-1}} \left(\frac{m \lambda e^{-g v}g}{h(g+\lambda)}\right)^k \frac{1}{k!}\right)
\end{equation}
where $u_n = 1$, if $n=0$, and $u_n=0$, otherwise. This gives us the probability distribution for the number of secondary bites that happen when there is a single infected human.

The power series in the equation converges to $a _1F_1(n+1,2,a)$, where $a=\left(\frac{m \lambda e^{-g v}g}{h(g+\lambda)}\right)$ and $_1F_1$ is the confluent hypergeometric function. Hence,
\begin{equation}
P(B_{SI}=n) = e^{-m \lambda \frac{e^{-g v}}{h}} \left( u_n + \left(\frac{\lambda}{g+\lambda}\right)^n \left(\frac{m \lambda e^{-g v}g}{h(g+\lambda)}\right) { }_1F_1(n+1,2,\left(\frac{m \lambda e^{-g v}g}{h(g+\lambda)}\right))\right)
\end{equation}





\begin{table}
\caption{Model: variables and their descriptions}

\begin{tabular}{ll}

$m$ & Total number of mosquitoes\\
$h$ & Total number of humans\\
$\lambda$ & biting rate \\
$g$ & parameter for the survival function, when exponentially distributed\\
$v$ & incubation time, when considered deterministic\\
$B$ & Number of individual bites in a period of time -- Poisson distributed\\
$B_T$ & Total number of bites in the period of time\\
$B_I$ & Total number of bites that turn mosquitoes into an virus--carrying state\\
$B_L$ & Number of bites of a mosquito during its period in virus--carrying state\\
$B_{SI}$ & Total number of bites by the group of virus--carrying mosquitoes\\
$N_S$ & Number of survivals of mosquitoes\\
$S$ & Survival time after biting an infected person\\
$V$ & Random variable for the incubation time\\
\end{tabular}

\end{table}














\section*{Discussion}

We believe this formulation can be important for studying dynamics of infectious diseases transmitted by vectors such as malaria, dengue and chikungunya.  

From our formula for the vectorial capacity process (moment--generating function), we are able to obtain all moments of the vectorial capacity process.
We demonstrate this in the case of simple incubation period and exponential distribution for survival.

It is desirable to collect data from experiments reported in the literature and possibly experiments designed to measure in the field the distributions that describe the biting process and survival process in order to construct generating functions that present good fitting to the observed data. 
We intend to further advance our research in these directions.


\bibliographystyle{abbrvnat}
\bibliography{vectorial.bib} 

\end{document}